# Predicting pigging operations in oil pipelines

Riccardo Angelo Giro[1, *], Giancarlo Bernasconi[1], Giuseppe Giunta[2], Simone Cesari[2]
[1] Politecnico di Milano, Milan
[2] Eni S.p.A., San Donato Milanese
* Corresponding author (email: riccardoangelo.giro@polimi.it)

## ABSTRACT

This paper presents an innovative machine learning methodology that leverages on long-term vibroacoustic measurements to perform automated predictions of the needed pigging operations in crude oil trunklines. Historical pressure signals have been collected by Eni (e-vpms® monitoring system) for two years on discrete points at a relative distance of 30-35 km along an oil pipeline (100 km length, 16" ID diameter pipes) located in Northern Italy. In order to speed up the activity and to check the operation logs, a tool has been implemented to automatically highlight the historical pig operations performed on the line. Such a tool is capable of detecting, in the observed pressure measurements, the acoustic noise generated by the travelling pig. All the data sets have been reanalyzed and exploited by using field data validations to guide a decision tree regressor (DTR). Several statistical indicators, computed from pressure head loss between line segments, are fed to the DTR, which automatically outputs probability values indicating the possible need for pigging the pipeline. The procedure is applied to the vibroacoustic signals of each pair of consecutive monitoring stations, such that the proposed predictive maintenance strategy is capable of tracking the conditions of individual pipeline sections, thus determining which portion of the conduit is subject to the highest occlusion levels in order to optimize the clean-up operations. Prediction accuracy is assessed by evaluating the typical metrics used in statistical analysis of regression problems, such as the Root Mean Squared Error (RMSE).

## 1. INTRODUCTION

Pipeline transportation systems represent the cheapest and safest solution to convey hydrocarbons, gases and other fluids over long distances. After construction, proper maintenance operations should be scheduled to preserve the integrity of such assets and to guarantee the desired transportation efficiency, since pipe internals naturally tend to accumulate deposits, such as rust, dirt, mill scale or paraffin wax [1]. Those constituents need to be removed for a number of reasons: firstly, to avoid product contamination, which can have a negative economic impact on the business; secondly, to allow for a better use of corrosion inhibitors, whose action is less effective if the pipe bore is covered with mill scale or partially corroded; thirdly, to improve flow rate and efficiency, which is maximized when the pipeline is completely clean (especially for pipelines having length of several tens of km); lastly, to facilitate pipeline drying, required to prevent internal corrosion and the formation of hydrates in natural gas transportation lines.

To perform internal cleaning of a pipeline, several techniques can be individually or jointly applied, such as: injection of chemical solvents, e.g., flux; internal sandblasting, in which an abrasive material is used to scrape the inner surface of pipe and to remove contaminants; purging with air or gas to prevent oxidation phenomena leading to corrosion; running a Pipeline Inspection Gauge (PIG), which is a multipurpose maintenance tool capable of flushing debris out of the pipe by scraping its internals with metallic brushes or plastic disks. Among those cleaning solutions, the PIG becomes particularly advantageous from a product saving and environmental point of view, especially in multi-product lines in which fluids are conveyed in batches: for instance, at the end of a given oil transfer, one can clear out the residuals stuck inside the pipe bore with a PIG run, thus allowing for a faster product switch. In addition, after each batch, one can avoid flushing the line with water, solvents or (in some cases) the following product: in such circumstances, fluids would undergo effluent treatment or contaminated product recovery; thanks to displacement pigging systems, these problems are easily overcome.

To this date, pipeline cleaning operations are still performed by following empirical rules or heuristics, since we lack a clear definition of clean pipeline and a rigorous method for measuring it. Currently available models in the literature [2, 3, 4, 5, 6] mainly focus on predicting wax deposition rate inside the pipes or have proven their effectiveness only on very short pipe sections (a few meters). This work, instead, presents an innovative machine learning methodology that makes use of pressure measurements, collected in discrete points along a crude oil pipeline, to automatically provide as output a numeric indicator that quantifies the cleanliness level of the pipeline itself, thus offering a clear indication of when a PIG campaign should be triggered. In addition, we demonstrate that our proposal can track the occlusion levels of individual pipe subsections, therefore determining which portion of the conduit is mostly blocked by wax deposits and debris.

The remainder of the paper is structured as follows. Section 2 outlines the experiment setup and the main data processing operations. Section 3 provides short- and long-term analyses of the head loss levels of the pipeline. Section 4 describes the design phase of the proposed data-driven pigging operations predictor. Section 5 presents the results obtained so far and Section 6 draws some conclusions.

## 2. EXPERIMENT SETUP AND DATA PROCESSING

This work makes use of historical pressure signals collected by a proprietary vibroacoustic monitoring system (e-vpms® technology [7, 8]), installed on a crude oil transportation line by connecting the Eni logistic terminals of Chivasso and Pollein, located in North Italy [9]. Such a line has a length of approximately 100 km and is characterized by 16" ID pipes. The satellite map of the conduit and the location of the recording stations are displayed in Fig. 1, respectively with a red line and yellow pins. Each e-vpms® measurement unit is equipped with a pressure transducer, recording the absolute pressure of the transported fluid, and a dynamic hydrophone, which measures small-scale dynamic pressure variations (in the order of kPa). Pressure data were collected at a sampling rate of 20 Hz, with acquisition interval of interest spanning from June $1^{st}$, 2013 up to December $1^{st}$, 2014. The distances between each station and the pumping equipment located at terminal A are reported in Table I.

Table I. Distance between each e-vpms station and the pumping terminal positioned in station A.

| Station | Distance with respect to station A (km) |
|---------|------------------------------------------|
| A | 0 |
| B | 59.307 |
| C | 100.486 |

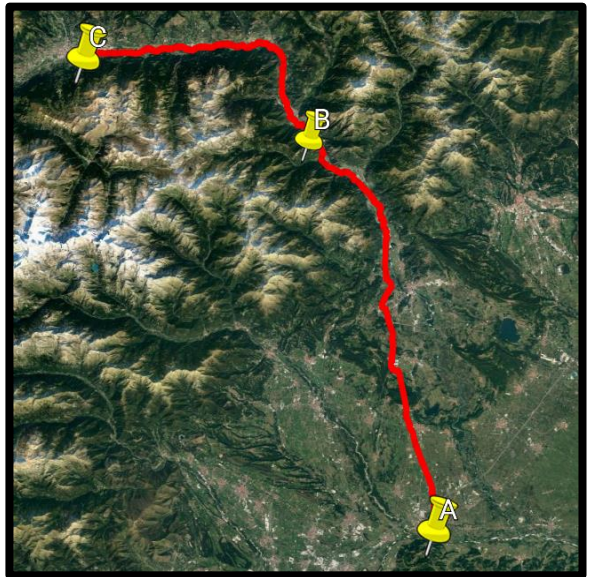

Fig. 1. Satellite map of Chivasso-Pollein pipeline long 100 km (red curve) and location of the e-vpms® measurement stations in North Italy (yellow pins).

The first processing step consists in transforming the unprocessed measurements into a format suitable for machine learning tasks. Fig. 2 shows the original, raw static pressure signals, collected from the three different e-vpms stations (A, B and C, respectively identified with turquoise, purple and green lines). Each time series needs to be cleansed to remove unwanted data points and compensate for the following undesired effects:

1) Presence of outliers due to sensor errors. In this specific case, static pressure readings lower than 0.5 bar and higher than 80 bar are discarded; likewise, dynamic pressure values below -200 kPa and above 200 kPa are eliminated from the dataset. These outliers are due to rare electromagnetic disturbances affecting the power unit of the measuring stations;

2) Unwanted pressure values corresponding to operational statuses of the line not contributing significantly to the occlusion levels of the pipes. More specifically, we assume that the formation of deposits within pipe segments mainly occurs when the oil is actively conveyed through the line: therefore, all the corresponding time intervals in which the pipeline is not operational (e.g., off) or is into a flow regulation state (e.g., pressure transients generated by pumping fluctuations) should be ruled out from the dataset;

3) Absolute pressure differences due to altitude variations between e-vpms stations. This adds a constant bias in static pressure measurements and is a function of both fluid density and altitude above mean sea level (a.m.s.l.).

Even though each of these impairments can be addressed manually, tackling the second issue by hand becomes impractical when processing datasets having billions of points. A possible solution to this problem consists in exploiting some automated detection procedure [10], such as the data-driven pump monitoring system described in [11]: we have therefore applied a Gaussian Mixture Model (GMM) based clustering algorithm to automatically retrieve, as output, a set of categorical labels, indicating all the time instants in which the system is either off or it is performing flow regulations. The corresponding datapoints can therefore be easily identified and removed, and the result of such an operation is displayed in Fig. 3.

The curves displayed in the chart have also been polished by all the outliers, as explained at the beginning of Section 2. Successively, the same set of operations has also been carried out on dynamic pressure measurements.

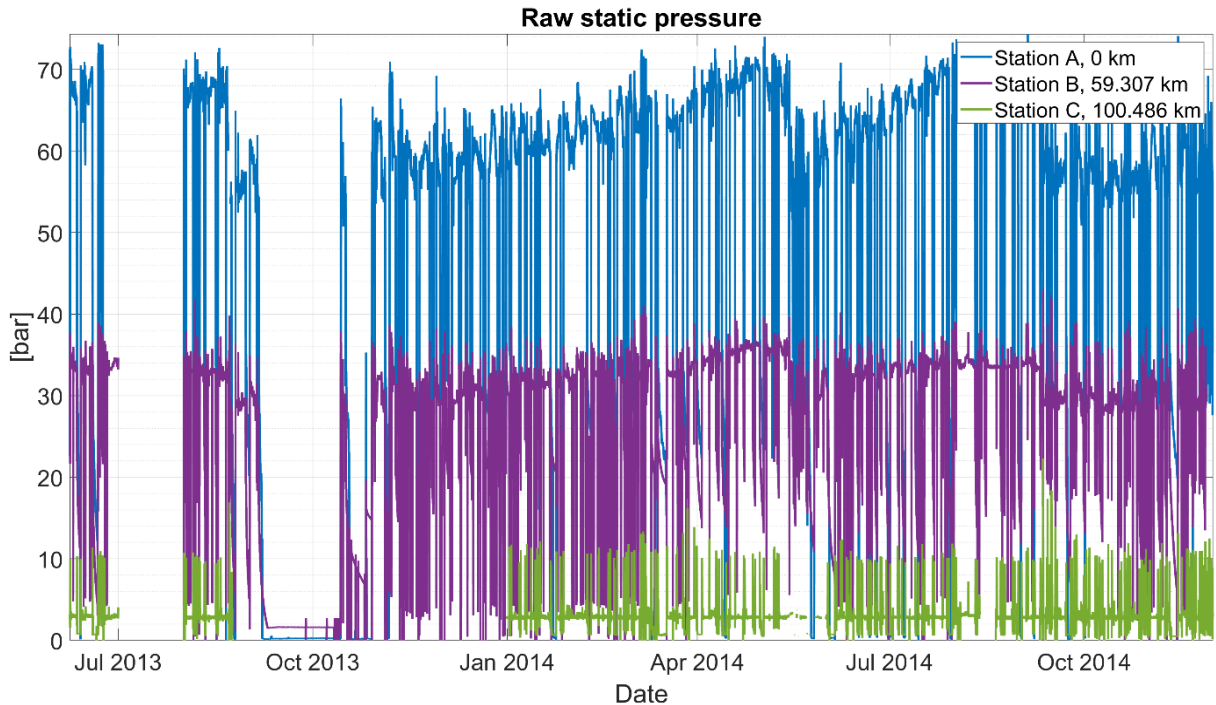

Fig. 2. Raw static pressure data.

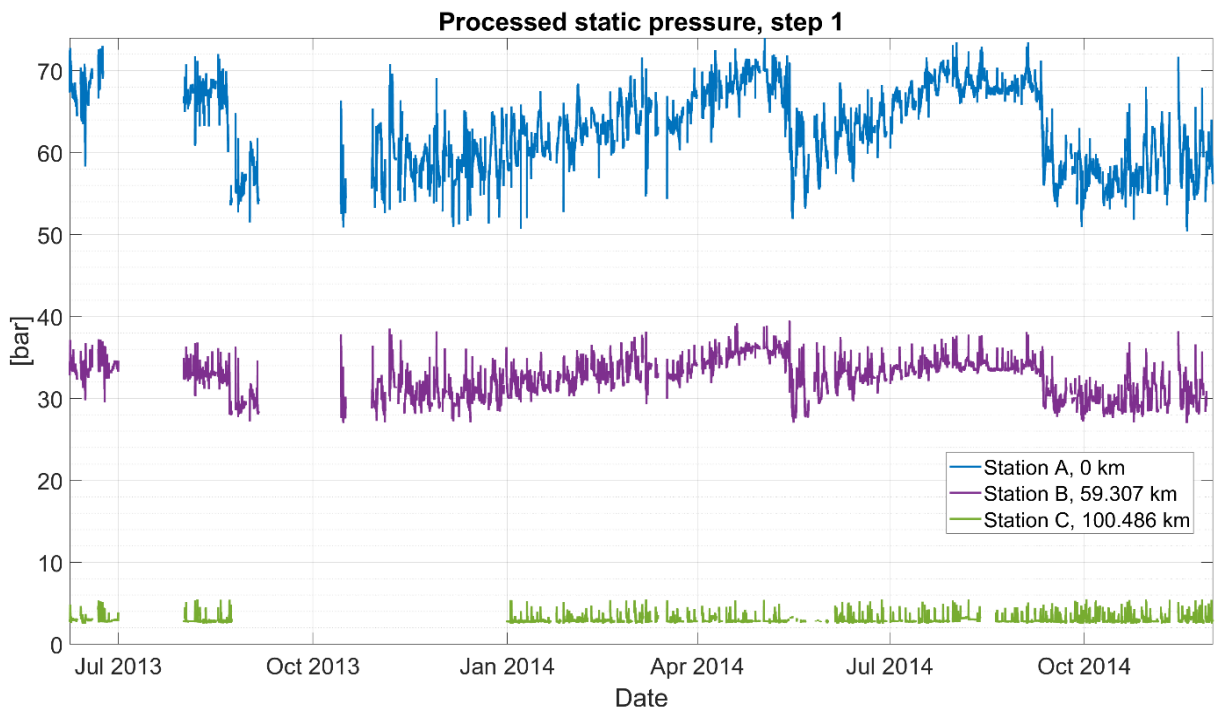

Fig. 3. Static pressure data after the first processing step.

The last processing step is only devoted to static pressure measurements. It is necessary to compensate for the differences in altitude a.m.s.l. between e-vpms® stations. Recalling that a pressure differential $dP$ can be expressed as a function of a relative altitude differential $dz$, one has that:

$$dP = -\rho \cdot g \cdot dz \ [\text{Pa}], \tag{1}$$

where $\rho$ is the density of a fluid (kg/m³) and $g$ is the gravitational acceleration ($\approx 9.81$ m/s²). The compensated pressure $P'$ will therefore be equal to:

$$P' = P - dP \ [\text{Pa}], \tag{2}$$

where $P$ is measured in Pa and corresponds to the original measurement. Lastly, recalling that 1 bar = $10^5$ Pa, one gets:

$$P' = \frac{P - dP}{10^5} \ [\text{bar}]. \tag{3}$$

Table II summarizes the values of $dP$ for stations B and C, with respect to A. Since no detailed information about the transported fluid was available, we have assumed an average density $\rho$ equal to 900 kg/m³, which corresponds to a typical medium-weight crude oil. Even though this approximation cannot faithfully represent the true scenario, in which the density of the transported fluid within the pipeline changes from day to day, it can still be considered acceptable: if we consider extreme values of $\rho$ (830 and 1000 kg/m³, respectively corresponding to very light and heavy crude oils), the error between the estimated and the true value of $dP$ would be (in our case) at most 3 bar. Lastly, Fig. 4 represents the result of this processing operation on the static pressure signals depicted in Fig. 3.

Table II. Pressure differentials due to altitude differences between e-vpms® stations.

| Station | Altitude a.m.s.l. (m) | $dz$ with respect to A (m) | $dP$ with respect to A (m) |
|---------|----------------------|----------------------------|----------------------------|
| **A** | 179 | 0 | 0 |
| **B** | 359 | 180 | -15.8922 |
| **C** | 558 | 379 | -33.4619 |

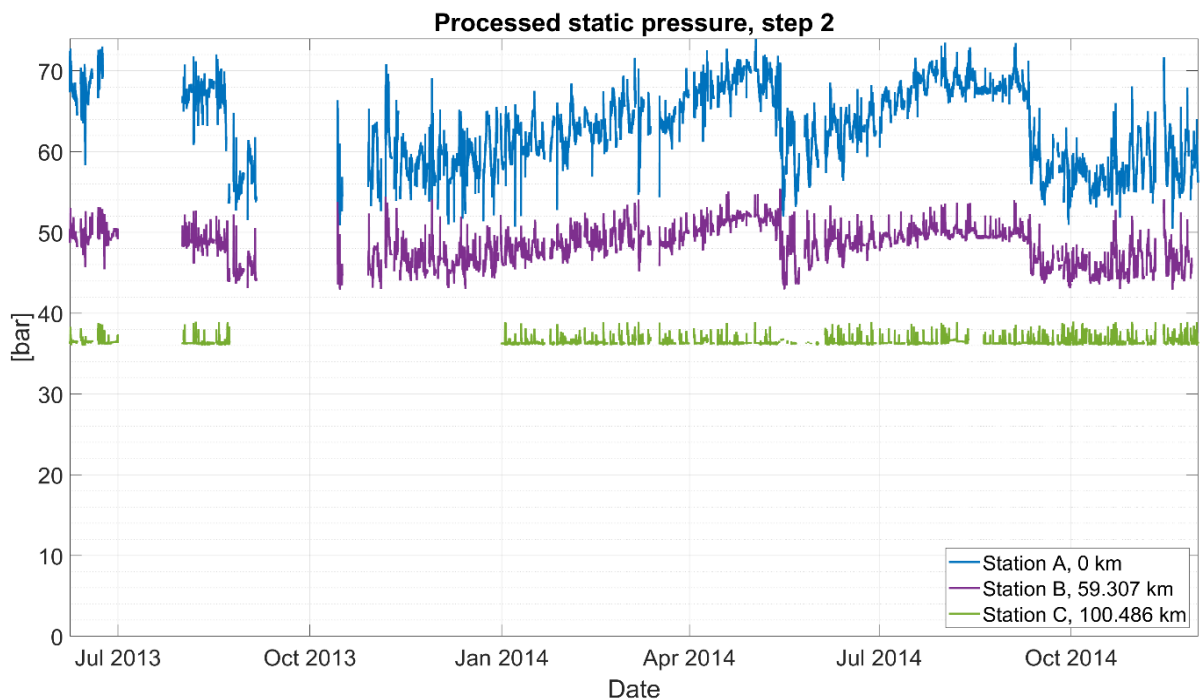

Fig. 4. Static pressure data after the second processing operation.

### 3.    HEAD LOSS ANALYSIS

The processed static pressure data displayed in Fig.  4 are successively employed to compute the head loss $h$ between line sections, representing the absolute pressure drop between a pair of e-vpms® stations [12]. Such measurements are then normalized with respect to the distance in km between the two locations, therefore $h$ has units of bar/km. Fig.  5 represents the evolution of the head loss, as a function of time, for three different line segments:

- $\overline{AC}$ (Fig.  5, top);
- $\overline{AB}$ (Fig.  5, middle);
- $\overline{BC}$ (Fig.  5, bottom);

Each chart displays the instantaneous value of $h$ (gray line) and is overlaid by its long-term trend, obtained by smoothing the former curve with a 1-week moving average (Fig.  5, from top to bottom: red, green and blue lines). We can observe that, in all three cases, the long-term head loss curves are characterized by a slow and gradual increase over the course of several weeks or months, coupled by rapid decreases having a much shorter temporal duration: the former phenomenon is mainly due to a progressive augmentation in the occlusion levels of the pipes; the latter correspond to the pigging campaigns performed on the pipeline, some of which have been highlighted in Fig.  6 using black vertical bars. It can be noted that every major drop in head loss occurs right after a pigging operation has been executed: in such circumstances, cleaner pipe sections allow the pumping terminal located at station A to operate with a lower service pressure while still delivering the reference value of about 36 bar at station C.

To validate the occurrence of each PIG campaign, we have developed a software tool capable of detecting, in the observed pressure measurements, the acoustic noise generated by the travelling PIG. An example of the output provided by such a software is displayed in Fig.  7, where the position of the PIG inside the pipeline has been tracked for one of the three main campaigns corresponding to the black vertical bars displayed in each chart of Fig.  6. If we consider Fig.  7, the latter represents a density plot of the normalized cross-correlation $R_{xy}$ between the dynamic pressure transients recorded by the hydrophones at stations A and B, as a function of time. Darker regions of the image correspond to values of $R_{xy}$ closer to 1 (maximum correlation); similarly, lighter areas present the lowest cross-correlation values, which tend to 0. At first sight, a horizontal dark line located at the cross-correlation time $\tau \approx -50$ s can be noticed: it corresponds to the physiological propagation delay of the acoustic waves emitted by the pump located at station A that reach station B.

Successively, on August 21st at approximately 08:00, a pipeline cleaning operation is initiated, as the PIG departs from station A to reach the end of the line (station C) about 24 hours later: a) such an event is observable in the cross-correlation map, as a slant dark line originates from $\tau \approx -50$ s and gradually increases towards the positive values of $\tau$ as time passes: b) this phenomenon can be interpreted as the position of an acoustic reflector (e.g., the PIG) that is travelling inside the pipes. In the same figure, a set of slant lines, parallel to the darkest one, can also be noticed: they are not representative of additional PIGs, but they are simply related to acoustic resonance effects generated by the moving inspection gauge [13].

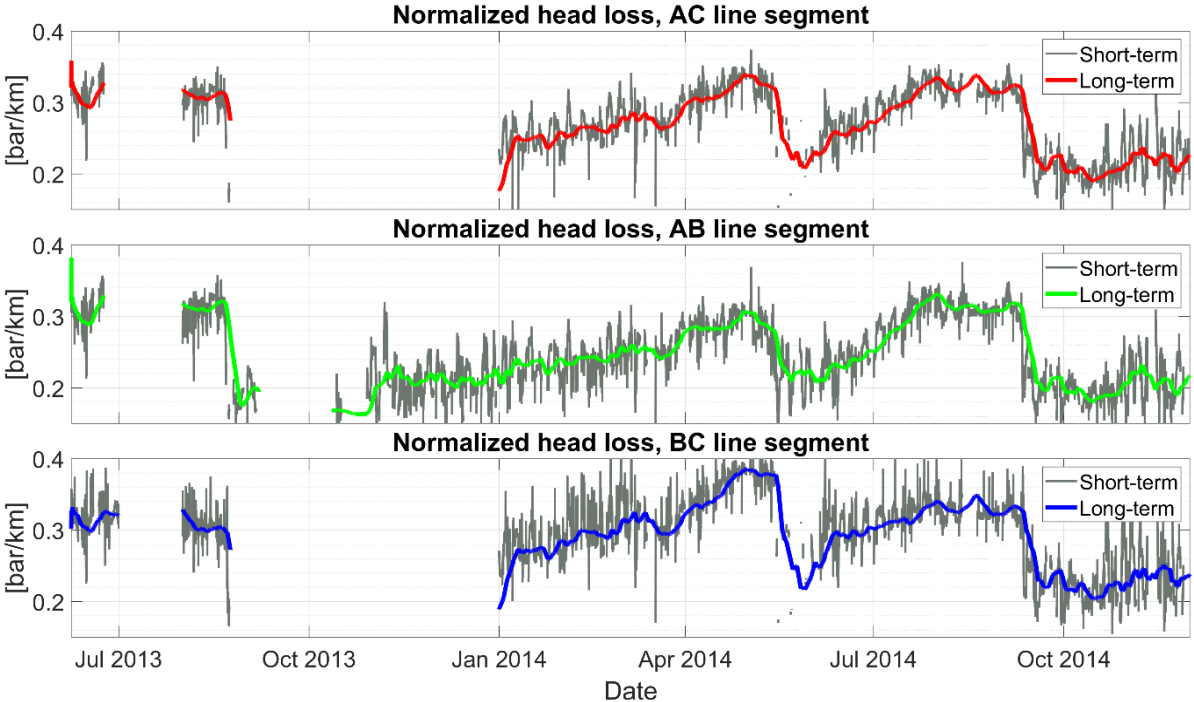

Fig. 5. Normalized head loss (short- and long-term values) for three different line sections.

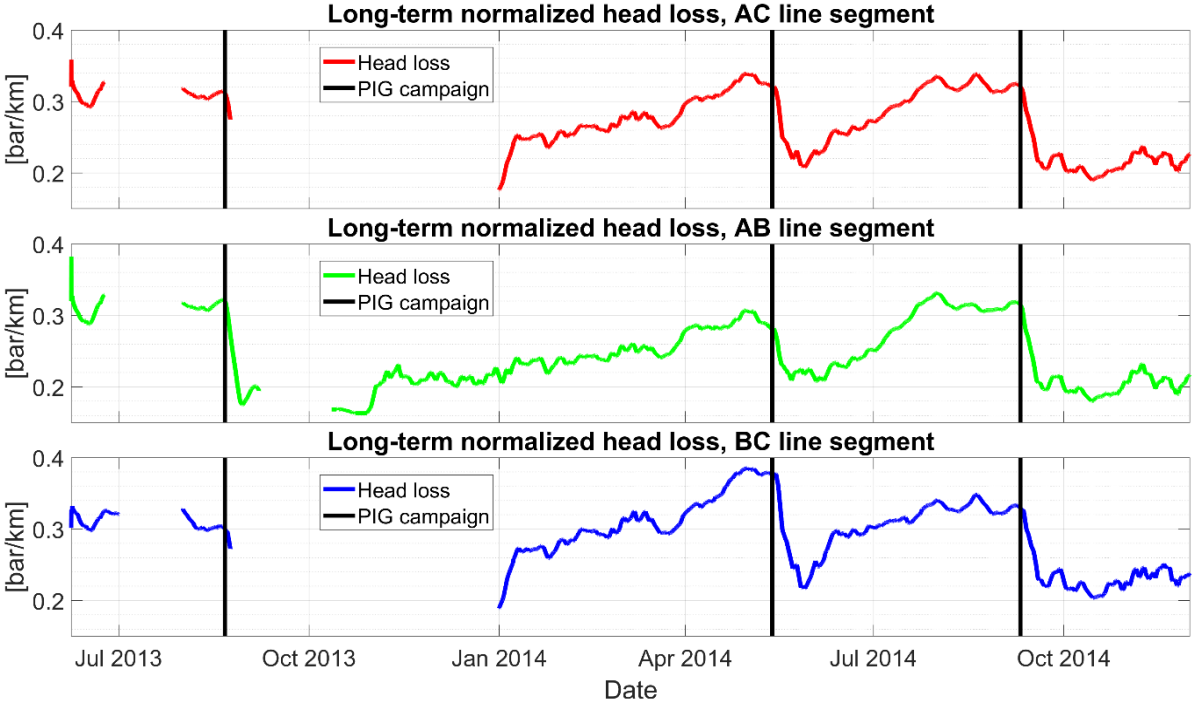

Fig. 6. Long-term normalized head loss (red, green and blue lines) for three different line sections and markers (black vertical lines) indicating the main PIG campaigns performed between 2013 and 2014.

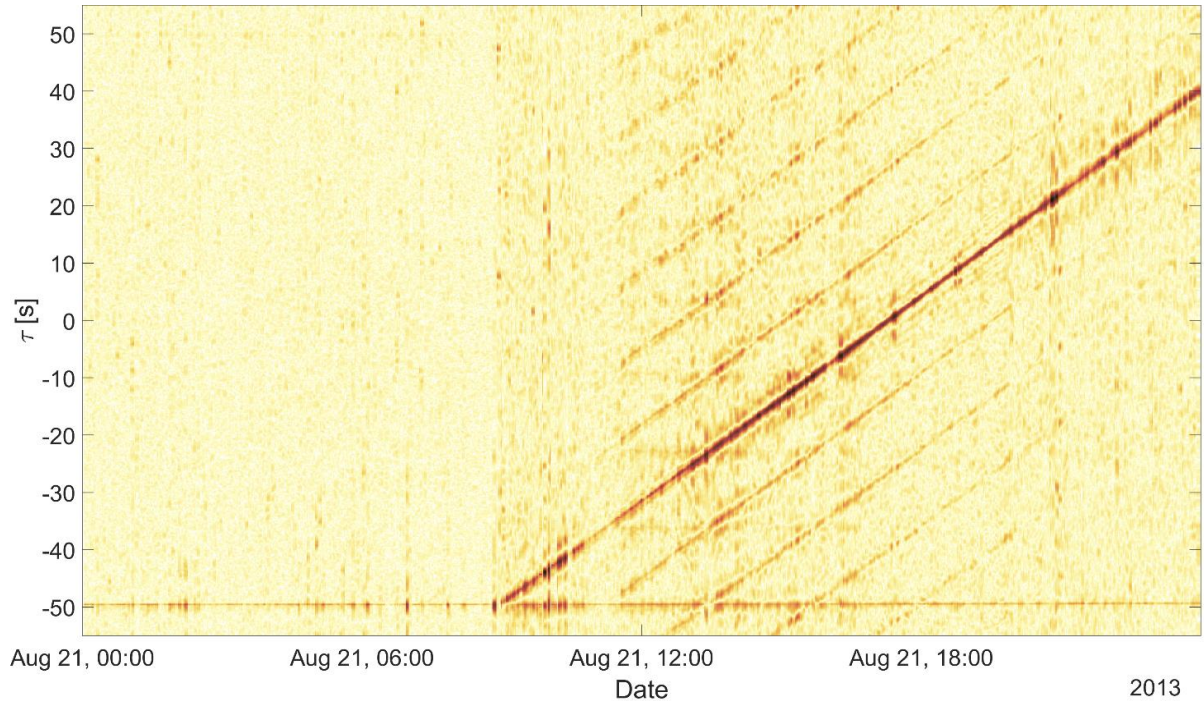

Fig. 7. PIG tracking on August 21st, 2013.

## 4. DATA-DRIVEN PIGGING OPERATIONS PREDICTOR

As stated in Section 1, the goal of this work consists in the development a data-driven procedure that, starting from the short-term head loss time series (gray curves in Fig. 5), can predict the long-term evolution of the occlusion level of Chivasso-Pollein pipeline segments. We have observed in Section 3 a strong correlation with the normalized head loss values; however, we still need to provide a clear and unambiguous definition that numerically quantifies the concept of pipe occlusion: this operation becomes necessary to entirely formalize the problem within a proper machine learning context.

To this purpose, a first step consists in defining which type of learning task needs to be solved; we have opted for supervised regression for two main reasons: firstly, pigging operations are performed as a consequence of multiple factors that continuously evolve over the course of several months (e.g., buildup of wax deposits, head loss variations); as a consequence, classification algorithms are not suitable for this kind of predictions, as they provide discrete outputs (e.g., binary labels, such as: clean pipe, clogged pipe); lastly, by predicting the value of a continuous variable through regression, we can express such a variable as a probability measure that is easily understandable by non-experts on the field. For instance, an automated system can be set up such that if the predicted pigging probability is above a certain threshold, a clean-up campaign is consequently triggered in the pipeline.

Employing supervised learning techniques requires having labelled data at disposal, which are rarely available in pipeline transportation systems [14]: to overcome this issue, we have manually built a target function $y$ (namely a PIG indicator) to be learned by the supervised regressor, and it corresponds to a pigging probability measure. In other words, the PIG indicator $y_{\overline{AB}}$ for a given pipeline segment $\overline{AB}$ between two stations $A$ and $B$ is defined as:

$$y_{\overline{AB}} = f(h_{\overline{AB}}), \qquad (4)$$

where $f$ is a non-linear mapping function of the long-term head loss $h_{\overline{AB}}$ between stations $A$ and $B$ (e.g., the green curve in Fig. 5) which transforms the data and rescales them in a range comprised between 0 and 1. The result of such an operation is displayed in Fig. 8, where the time series of several PIG indicators, corresponding to the three-line sections listed at the beginning of Section 3, are represented. Each curve presents some gaps, which correspond to missing head loss data: in all those circumstances the machine learning algorithm cannot either trained nor tested.

The learning task is performed by a Decision Tree Regressor (DTR), which is a supervised, data-driven meta-estimator trained to automatically provide pigging probability values by analyzing certain characteristics (features) of one or more given input signals. The working principle behind DTRs consists in training a machine to predict the values a continuous random variable based on a set of simple and intuitive decision rules, which are directly derived from the input features. The latter consist in a set of statistical indicators, each computed over rolling windows of 8, 16 and 24 hours: more precisely, we have chosen to evaluate the mean, minimum and maximum values of the short-term head loss, thus resulting in a final set of 9 features.

The DTR model has been trained using data from $\overline{AC}$ line segment (from June 1st, 2013 to May 31st, 2014), whereas the testing phase is firstly performed on the same line section (from June 1st, 2014 to December 1st, 2014) and successively on the remaining ones ($\overline{AB}$ and $\overline{BC}$, both from June 1st, 2013 to December 1st, 2014). This threefold subdivision on the test set has two main purposes: firstly, evaluate the model capabilities on data belonging to the same distribution of the training set; lastly, test measurements collected in different line sections to determine which portion of the conduit is subject to the highest occlusion levels among the others.

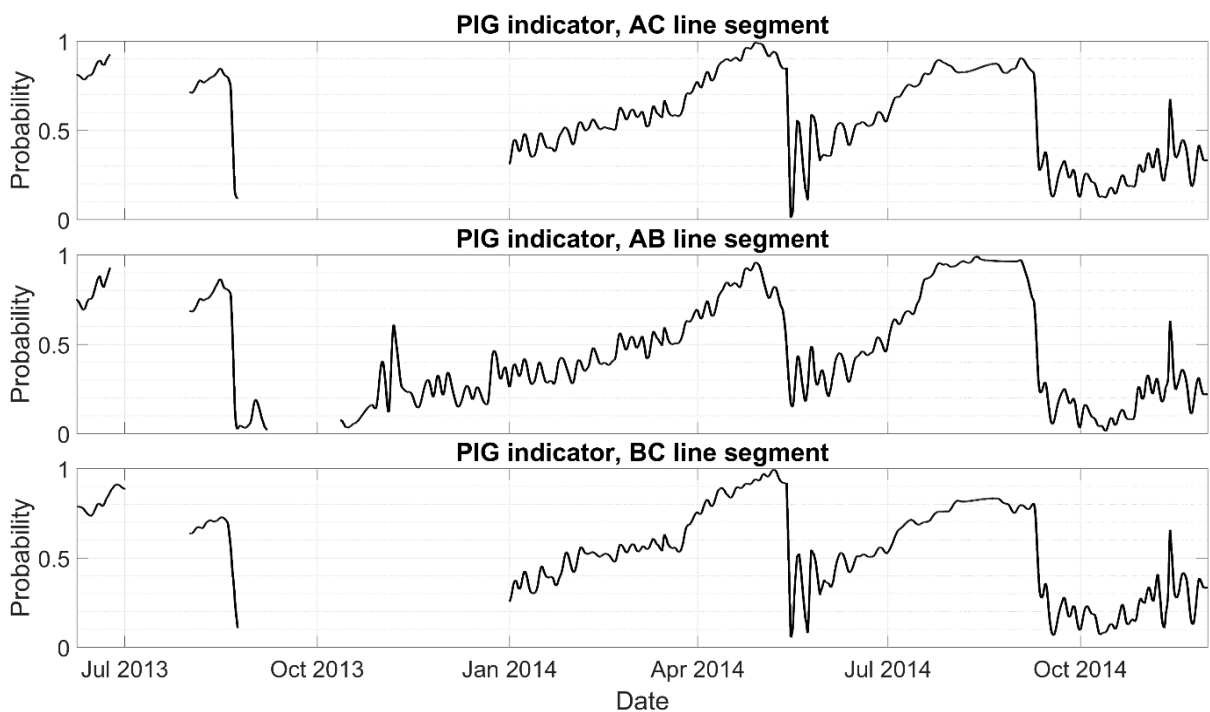

Fig. 8. PIG indicators for three different line sections.

## 5. RESULTS

The accuracy of the model has been assessed by evaluating, for each line segment, the Root Mean Squared (RMS) values of the estimation error $\epsilon$ between the estimated pigging probability $\hat{y}$ and the target PIG indicator $y$. $\epsilon$ is defined as follows:

$$\epsilon = \hat{y} - y. \qquad (5)$$

Since the target function $y$ had been transformed to be bound between 0 and 1, one can also derive the percentage prediction accuracy as $100 \cdot (1 - RMS\{\epsilon\})$. Table III reports the values of RMS$\{\epsilon\}$ and of the corresponding prediction accuracy for the three pipeline segments listed in Section 3, while Fig. 9 graphically compares the true (black line) and the predicted (red curves) values of the PIG indicators. The results are quite satisfactory, as the attained accuracy level is greater than 97% in all three line section.

Table III. RMS values of the estimation error and corresponding prediction accuracy.

| Line section | $RMS\{\epsilon\}$ | Accuracy |
|---|---|---|
| $\overline{AC}$ | 0.0274 | 97.26% |
| $\overline{AB}$ | 0.0262 | 97.38% |
| $\overline{BC}$ | 0.0244 | 97.56% |

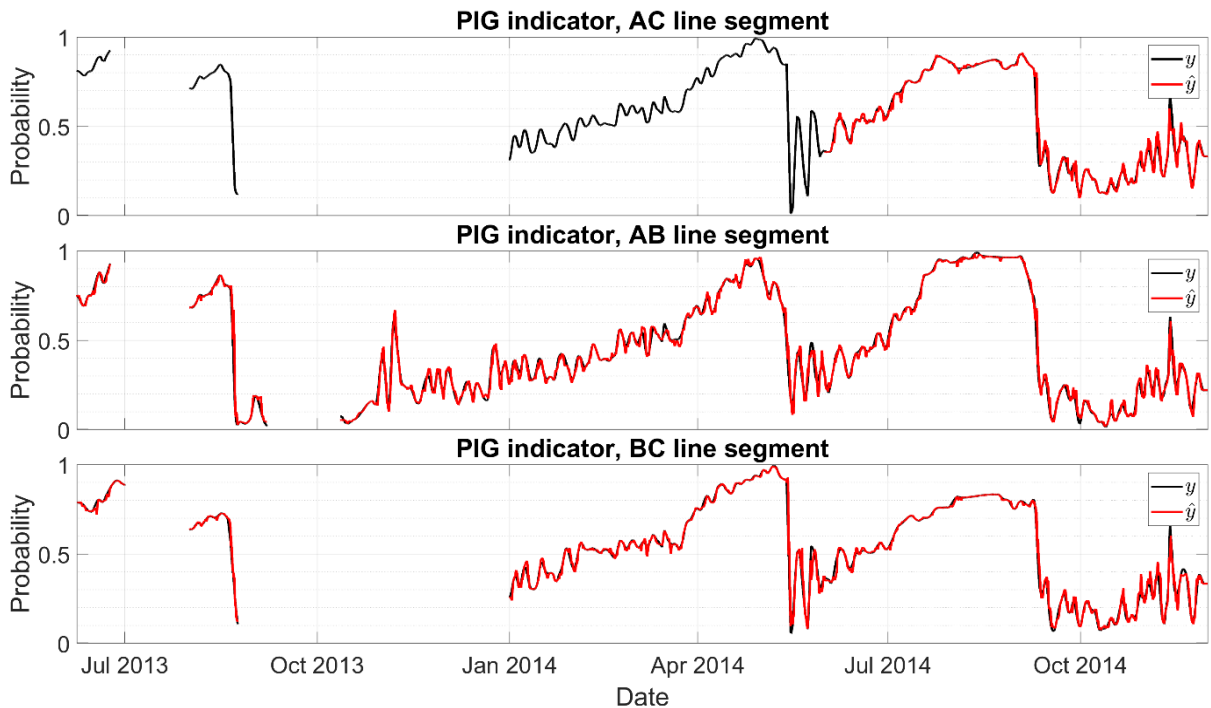

Fig. 9. Ground truth (black lines) and predicted (red lines) values of the PIG indicators for three different line segments.

## 6. CONCLUSIONS

A data-driven methodology to perform automated predictions of the needed pigging operations in crude oil pipelines has been presented. The proposed solution makes use of standard pressure measurements, collected in at least two different locations along the pipeline, which are reprocessed and fed to a non-linear, supervised regression algorithm (DTR). The latter has been designed to provide as output a set of probability measures (PIG indicators) which numerically quantify the necessity of performing a PIG campaign. Results obtained so far show the possibility of predicting and tracking the occlusion levels of the entire pipeline and of individual pipe sections: such capabilities prove to be advantageous in the context of planning optimal predictive maintenance strategies, as PIG campaigns can be triggered only when necessary and on the mostly clogged pipeline sections.

Performance has been assessed in terms of prediction accuracy, obtaining an average score greater than 97% for the three tests analyzed. Future work includes an additional validation phase on other crude oil transportation systems, testing on multiphase pipelines in upstream scenario and the definition of optimal threshold criteria which would trigger a new PIG campaign.

## ACKNOWLEDGMENT

This research was mainly carried out in the framework of the R&D – DIONISIO project founded by Eni S.p.A. The authors are grateful to Eni Logistic Department and SolAres JV teams for technical support during the field tests.